\documentclass[12pt]{iopart}
\usepackage{iopams}  
\usepackage{graphicx}
\usepackage{ulem}
\usepackage[svgnames]{xcolor}
\usepackage{bm}
\usepackage{multirow}

\begin{document}

\title{Nernst response, viscosity and  mobile entropy in vortex liquids}

\author{Kamran Behnia}

\address{Laboratoire de Physique et Etude des Mat\'{e}riaux (CNRS- Sorbonne Universit\'e) \\ 
ESPCI Paris, PSL  University, 75005 Paris, France}
\ead{kamran.behnia@espci.fr}
\vspace{10pt}
\begin{indented}
\item[]August 2022
\end{indented}

\begin{abstract}
In a liquid of superconducting vortices, a longitudinal thermal gradient generates a transverse electric field. This Nernst signal peaks at an intermediate temperature and magnetic field, presumably where the entropy difference between the vortex core and the superfluid environment is largest. There is a puzzling similarity of the amplitude of this peak across many different superconductors. This peak can be assimilated to a minimum in the viscosity to entropy density ratio of the vortex liquid. Expressed in units of $\frac{\hbar}{k_B}$, this minimum is one order of magnitude larger than what is seen in common liquids. Moreover, the entropy stocked in the vortex core is \textit{not} identical to the entropy bound to a moving magnetic flux line. Due to a steady exchange of normal quasi-particles, entropy can leak from the vortex core. A slowly moving vortex will be peeled off its entropy within a distance of the order of a superconducting coherence length, provided that the $\frac{\Delta}{E_F}$ ratio is sufficiently large.

\end{abstract}

%
%
%
%
%

\section{Introduction}
In presence of a magnetic field, the thermoelectric response acquires an off-diagonal component. This is the Nernst coefficient, which is particularly large in dilute metals with highly mobile carriers~\cite{Behnia2016}. An emblematic case is elemental bismuth. The peak Nernst signal in (large and dislocation free crystals of) bismuth approaches 1 V/K at liquid He temperature and a magnetic field of 1 T~\cite{Galev1981}. Expressed in the natural unit for thermoelectric response ($k_B/e= 8.6 \times 10^{-5} V/K$), this becomes a very large number. Often, the Nernst response is much weaker than this natural unit. However, even when it is small as nV/K, it can be safely measured by many experimentalists.

The focus of the present paper is the Nernst signal generated by the vortex motion in a type II superconductor. This phenomenon was discovered in 1960s~\cite{LOWELL1967,solomon1967,HUEBENER1967947} and became the subject of several contemporaneous theoretical investigations \cite{Stephen1966,Clem1968,Maki1968}. A few decades later, it was also observed in superconducting cuprates \cite{Huebener_1995}, which host a remarkably broad vortex liquid state \cite{Blatter1994}.

Despite this long history, a recent experimental study by Rischau \textit{et al.}~\cite{Rischau2021} demonstrated that the vortex Nernst signal cannot be quantitatively understood with available theories. This is in surprising contrast with the case of the  Nernst signal generated by the superconducting fluctuations above the critical temperature. In the latter case, the theory originally formulated by Ussishkin, Sondhi and Huse \cite{Ussishkin} has proven to be a great success. It was confirmed and extended by other theorists~\cite{Serbyn2009,Michaeli2009,Levchenko2011} and was experimentally confirmed in both conventional \cite{Pourret2006,Pourret2007,Spathis2008} and cuprate \cite{kokanovic2009,Chang2012,Tafti2014} superconductors (For reviews, see \cite{Pourret2009,Behnia2016,Cyr2018}). 

Ironically, the exploration of the Nernst effect in this century was initially driven by attributing to vortex-like excitations  an unexplained Nernst signal  above the critical temperature in undoped cuprates \cite{Xu2000}. Thanks to numerous experiments which followed this report,  we have nowadays a reasonable understanding of the amplitude of the quasi-particle Nernst signal in metals and the fluctuating Nernst signal of in the normal state of superconductors \cite{Behnia2016}. The rough amplitude of the anomalous Nernst signal of magnets can also be guessed from their anomalous Hall conductivity \cite{Xu2020}. In contrast, we lack an established quantitative account of how mobile vortices give rise to a Nernst response.

The present paper aims to summarize what is known and what is yet to be understood about the measured vortex liquid Nernst signal in various superconductors. It identifies a link between an upper bound to the amplitude of the Nernst signal and a lower bound to the viscosity of the vortex liquid and highlights a distinction between two concepts of entropy of a superconducting vortex: the one statically stocked in its core and the one dynamically carried by the mobile flux line.

\section{The amplitude of S$_{xy}$ and its surprising invariability among superconductors}
\begin{figure*}
\centering
\includegraphics[width=16cm]{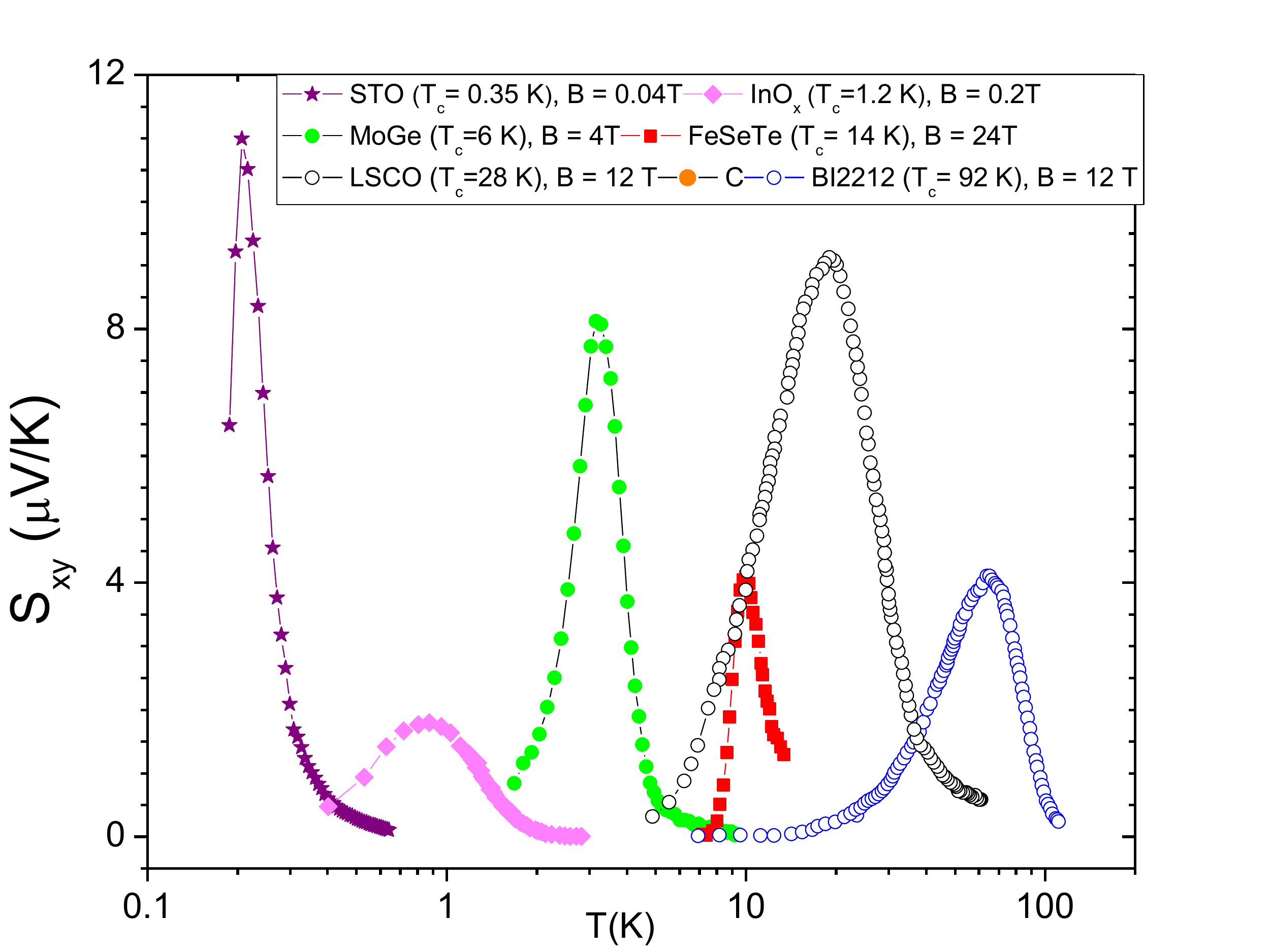}
\caption{\textbf{The amplitude of the Nernst signal in various superconductors:} Temperature dependence of the largest Nernst signal in several superconductors. Despite three orders of magnitude variation in their critical temperature, these superconductors display a roughly comparable vortex Nernst signal, which ranges between 2 and 11 $\mu V/K$. In contrast, the quasi-particle Nernst signal in metals can be many orders of magnitude larger ($ \approx mV/ K$) or many orders of magnitude smaller ($ \approx nV/ K$), according to the mobility and the Fermi energy of the metal in question. }
\label{fig:var-sup}
\end{figure*}
Experimentally, Nernst effect is measured by applying a thermal gradient along the crystal (let us call it the x-axis) and measuring an electric field in the lateral direction (the y-axis) in presence of a magnetic field perpendicular to both (along the z-axis). The Nernst signal is defined as: 

\begin{equation}
 S_{xy} =\frac{E_y}{\nabla_x T}
\label{Sxy}
\end{equation}

A striking fact about the amplitude of S$_{xy}$ in superconducting crystals or thin films measured in the past half a century is that it peaks to a value in the range of $\mu V/K$. This feature, first highlighted in ref. \cite{Rischau2021}, is illustrated in Fig.\ref{Sxy}. Despite the fact that the superconducting critical temperature is $\approx 0.3 K$ in strontium titanate  and $\approx 100 K$ in Bi2212, their Nernst peaks, occurring at widely different temperatures and magnetic field, are comparable in magnitude.  This is to be contrasted with the magnitude of  S$_{xy}$ in metals, where it can vary by many orders of magnitude \cite{Behnia2009,Behnia2016}. The S$_{xy}$ peak in bismuth, a dilute semi-metal hosting high-mobility carriers, is five to eight orders of magnitude larger \cite{Sugihara,Mangez1976,Behnia2007,Galev1981} than in dense metallic perovskytes \cite{Xu2008,Jin2021}. 

Let us examine a possible link between this observation and a general property of liquids.  

\subsection{A lower bound to viscosity of liquids}
The shear viscosity of a fluid, $\eta$, quantifies the tangential stress, $\sigma_{xy}$, required to generate a velocity gradient in the fluid \cite{falkovich2011fluid}:

\begin{equation}
\sigma_{xy} =\eta \frac{dv}{dz}
\label{eta}
\end{equation}

$\eta$ is the dynamic viscosity, which is expressed in units of Poise ($\equiv$ Pa.s). Divided by the particle mass, $m$ and particle density, $n$ of the fluid, it yields the kinematic viscosity, $\nu=\frac{\eta}{m n}$. The latter is equivalent to momentum diffusivity.

Note that the sign of viscosity can only be positive. It also happens that its amplitude cannot be arbitrarily small \cite{Purcell,Trachenko2021}. Indeed, the magnitude of kinematic viscosity is bounded by fundamental constants \cite{Trachenko2021}.  Trachenko and Brazhkin have demonstrated that, while the amplitude of viscosity of a fluid depends on the microscopic interactions among its constituent particles, it always exceeds a minimal bound, set by $m$ and the electron mass, $m_e$ \cite{Trachenko2020}: 

\begin{equation}
\nu_{min} = \frac{1}{4 \pi}\frac{\hbar}{\sqrt{m_e m}}
\label{nu}
\end{equation}

The temperature dependence of viscosity in common fluids displays a minimum, which approaches (while exceeding) $\nu_{min}$. The minimum arises because of the competition between a liquid-like regime (where warming diminishes viscosity by enabling particles to hop across energy barriers) and a gas-like regime (where warming strengthens viscosity by shortening their mean-free-path) \cite{Trachenko2020}. In supercritical fluids, this minimum in momentum diffusivity is accompanied by a similar minimum in thermal diffusivity \cite{Trachenko2021c}. 

This minimum has attracted recent attention in the context of the debate on a possible bound to the ratio of viscosity to entropy density in fluids \cite{Cremonini2011}. This gives us an opportunity to compare vortex liquids with common liquids.

\subsection{The viscosity-to-entropy density ratio}
Kinematic viscosity, $\eta$, can be expressed in units of Planck constant per volume and entropy density, $s$, in units of Boltzmann constant per volume. Therefore, their ratio, $\frac{\eta}{s}$, can be expressed in units of $\frac{\hbar}{k_B}$ for any fluid  It has been noticed that in common fluids this ratio approaches unity, but remains  above it (see Fig. \ref{fig:eta-min}.a) \cite{Cremonini2012}. This observation was made in the context of the debate following the discovery of a theoretical holographic bound to $\frac{\eta}{s}$ \cite{Policastro2001,Kovtun2005}.

Eq. \ref{nu} defines an explicit  minimum for kinematic viscosity. However, one can see that its also implies  a  minimum in $\frac{\eta}{s}$. Within a numerical factor of the order of unity, the entropy density of a classical liquid is of the order of $k_B$ times its molar density. The dynamic viscosity ($\eta$)  is the kinematic viscosity ($\nu$) times the mass density of the liquid and the particle mass ($m$) can be expressed in terms of the proton mass $m_P$. Therefore,  Eq. \ref{nu} leads to:

\begin{equation}
\frac{\eta}{s} \geq \frac{1}{4 \pi}\sqrt{\frac{m_p}{m_e}}\frac{\hbar}{k_B}
\label{etas}
\end{equation}

Here, $\frac{m_p}{m_e}= 1836$ is the proton-electron mass ratio. Thus, this inequality accounts for what is seen in Fig. \ref{fig:eta-min}.a.

\subsection{The  peak in S$_{xy}$ as a minimum in $\frac{\eta}{s}$}
Let us come back to the vortex liquid. The ratio of the quantum of the magnetic flux $\Phi_0=\frac{h}{2e}$ to the Nernst signal $S_{xy}$ can be expressed in units of $\frac{\hbar}{k_B}$. Let us show that it can be assimilated to the ratio of viscosity to the entropy density of the vortex liquid. 
\begin{figure*}
\centering
\includegraphics[width=14cm]{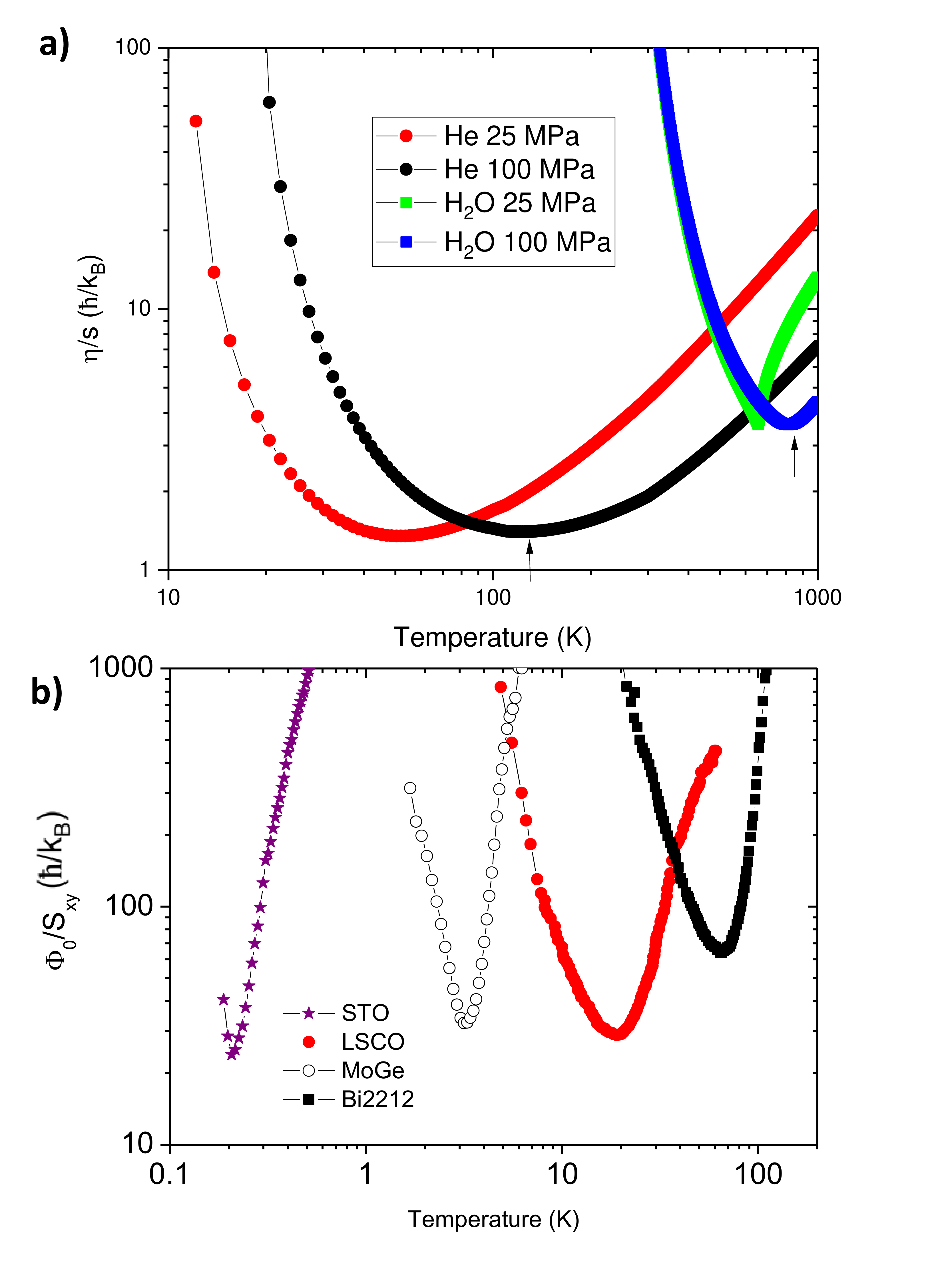}
\caption{\textbf{The viscosity-to-entropy density ratio:} a) The ratio of dynamic viscosity, $\eta$, to entropy density, s, in two fluids (H$_2$O and helium) at two different pressures (25 and 100 MPa). The data was extracted from the National Institute of Science and Technology (NIST) reference database \cite{nist}. Note the presence of a minimum, which is larger, but of the order of $\hbar/k_B$.  b) The ratio of the quantum of the magnetic flux  ($\Phi_0=\frac{h}{2e}$) to the Nernst signal ($S_{xy}=\frac{E_y}{\nabla_x}$) in several superconductors. This quantity is equivalent to viscosity to entropy density ratio of a vortex liquid (see text). Expressed in units of $\hbar/k_B$, it has a minimum, which is an order of magnitude larger than what is seen in classical fluids.}
\label{fig:eta-min}
\end{figure*}

The vortex Nernst signal arises due to the drift velocity of magnetic flux lines, $v_x$. Taking the number of vortices per unit area to be $n_V$, the Josephson equation yields the Electric field:
\begin{equation}
E_y=   n_V \Phi_0 v_x
\label{Josephson}
\end{equation}
This drift velocity along the x-axis is caused by the thermal force exerted by a thermal gradient, which is itself balanced by a viscous force:

\begin{equation}
\nabla_x T  s =  n_V \eta v_x
\label{n_vs}
\end{equation}

Compare this equation with eq. \ref{eta}, to note that the assumption here is that $\eta$ is a genuine shear velocity and the drift velocity of vortices can be assimilated to a gradient in local fluid velocity generated by thermal stress. Combining these two equations leads to the following expression for the Nernst signal: 

\begin{equation}
S_{xy}\equiv\frac{E_y}{\nabla_x T}= \Phi_0\frac{s}{\eta}
\label{N eta s}
\end{equation}

Therefore, the ratio of the Nernst signal to the quantum of the magnetic flux in a given vortex liquid equals the inverse of the ratio of the viscosity to the entropy density. 

Plotting $\frac{\Phi_0}{S_{xy}}$ for a few superconductors among those shown in Fig. \ref{fig:var-sup}, we can see that the viscosity-to-entropy-density  in different vortex liquids presents a minimum of comparable amplitude, which in units of  $\hbar/k_B$, exceeds what is seen in common liquids by one order of magnitude (Fig. \ref{fig:eta-min}b). 

Note the difference in the origin of minimum in the two cases. In common liquids, the non-monotonic temperature dependence is due to viscosity, in the vortex liquids, it is mostly due to the difference in entropy between the two components of the fluid. Nevertheless, the comparison suggests that the observed similarity of the amplitude of S$_{xy}$ across superconductors may be a consequence of a bound, set by fundamental constants, to the ratio of viscosity to entropy density in any liquid of superconducting vortices.
\section{ Extracting the  mobile entropy of vortices from $\alpha_{xy}$} 
\begin{figure*}
\centering
\includegraphics[width=16cm]{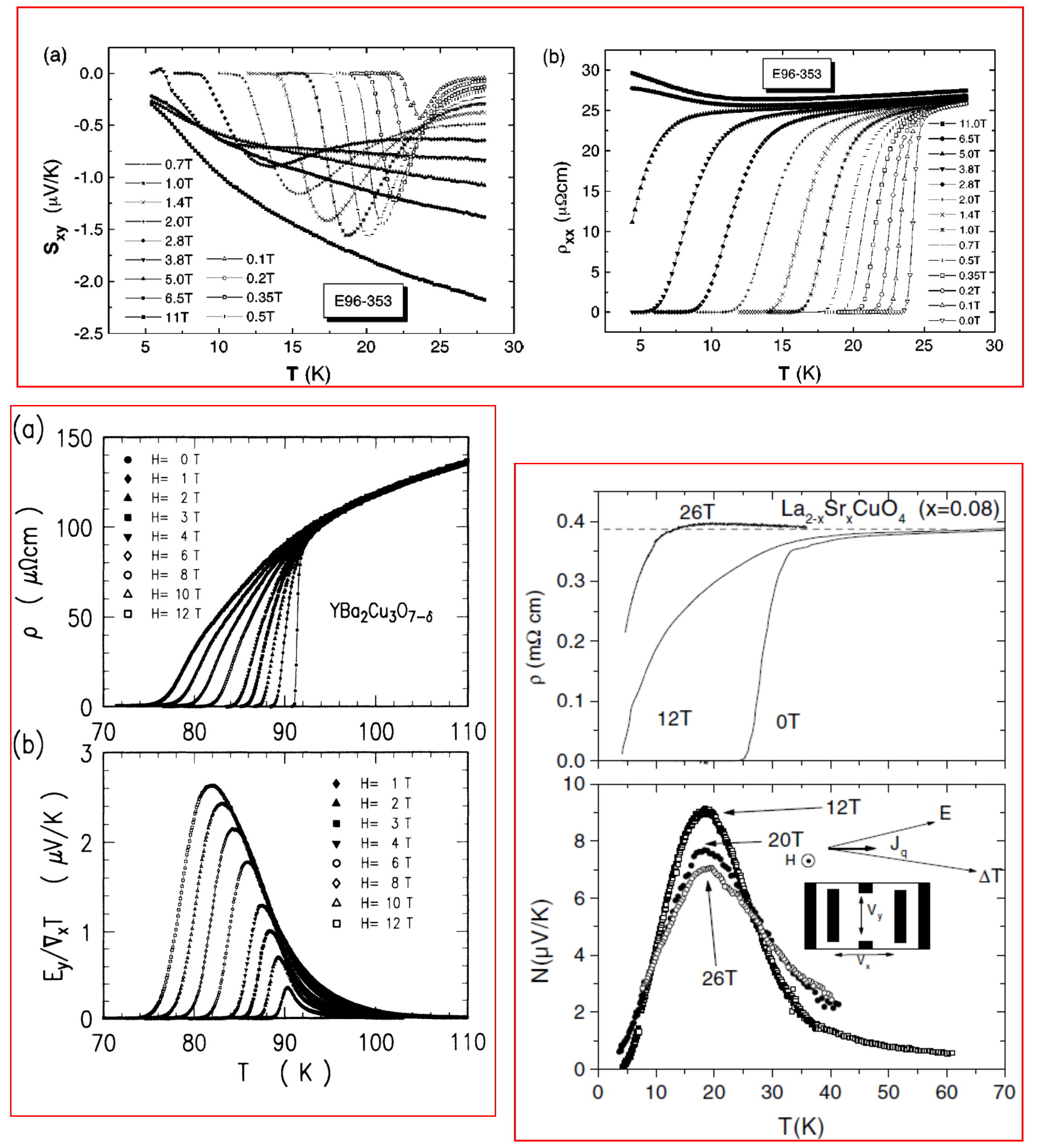}
\caption{\textbf{Nernst effect and resistivity in cuprates:} Published figures of the Nernst and resistivity data in three cuprate superconductors: Top  Nd$_{2-x}$Ce$_x$CuO$_4$ (NCCO) \cite{Gollnik1998}; Bottom left: YBa$_2$Cu$_3$O$_{7-\delta}$ (YBCO) \cite{Li1994}; La$_{1.92}$Sr$_{0.08}$CuO$_4$(LSCO) \cite{Capan2002} . Reprinted with permission; Copyright (1994, 1998, 2002) by the American Physical Society. In the  case of NCCO, the negative sign of the Nernst coefficient is due to the choice of a different sign convention.}
\label{fig:cuprates}
\end{figure*}

While the Nernst signal, S$_{xy}$ is what is immediately measured in a  Nernst experiment, more extensive attention has been given to $\alpha_{xy}$,  sometimes called the Nernst conductivity. It is the off-diagonal component of the thermoelectric conductivity tensor defined by $\overline{\rho}\overline{\alpha}= \overline{S}$, In general one has:

\begin{equation}
S_{xy}=\rho_{xx}\alpha_{xy}+\rho_{xy}\alpha_{xx}
\end{equation}

However, in our context, the second term can be safely neglected in  most cases. Then $\alpha_{xy}$  can be extracted from the measured S$_{xy}$ and the measured $\rho_{xx}$.  Note that in three dimensions, where resistivity is expressed in $\Omega$.m, the units of $\alpha_{xy}$ is A/K.m. But the length disappears in two dimensions, and $\alpha_{xy}$ can be expressed in a natural unit: $\frac{ek_B}{\hbar}\approx 21 nA/K$ \cite{Pourret2006}.  

In the flux flow regime, electrical resistivity is set by the balance between the Lorentz force and a viscous force. Therefore:  

\begin{equation}
\rho_{xx}=\Phi_0 \frac{n_V}{\eta}
\label{fFR}
\end{equation}

By assuming that the viscous force on vortices is the same in a flux flow experiment and in a Nernst experiment, one can link the amplitude of $\alpha_{xy}$ to the mobile entropy of a vortex per unit of length, $S_d$ \cite{Huebener1979}: 

\begin{equation}
\alpha_{xy}= \frac{S_{xy}}{\rho_{xx}}=\frac{S_d}{\Phi_0}
\end{equation}
In three dimensions, $S_d$ is expressed in units of $J.K^{-1}.m^{-1}$. Multiplying this quantity by a length leads to $S^d_{sheet}$, a quantity which can be expressed in units of Boltzmann constant. 

\subsection{Experimentally observed $\alpha_{xy}$ in superconductors}
Rischau \textit{et al.} observed that the peak $\alpha_{xy}$, extracted by dividing the maximum $S_{xy}$ in the vortex state to the resistivity at the same temperature and magnetic field reveals a general, and yet to be understood, trend. Here is a brief review of this point.

\textbf{Cuprates} - Numerous studies have been devoted to quantifying either the Ettingshausen ~\cite{Palstra1990} and the Nernst ~\cite{Li1994,Gollnik1998,Wang2001,Wang2002,Capan2002,Wang2006,Rullier2006,Balci2003} effects in the vortex state of the cuprates. Fig. \ref{fig:cuprates} reproduces the data from three different studies, where both the Nernst signal and the resistivity were reported. As seen in Table \ref{Tab1}, which summarizes the situation for four different cuprates, if one takes the c-axis lattice parameter to pass from $S^d$ to $S^d_{sheet}$, one finds that, its magnitude varies between 0.7 $k_B$ in LSCO to 4 $k_B$ in Bi2212. Boltzmann constants. In other words,  $S^d_{sheet}$ is of the order of magnitude of the Boltzmann constant, despite the fact that at each CuO$_2$ plane, thousands of quasi-particles are associated with the normal core of a vortex.

Note that the relatively large value of $S^d_{sheet}$ in Bi2212 is a consequence of its relatively longer c-axis parameter. Instead of the latter, one may  take the distance between copper-oxygen planes  (of the order of 0.6 to 0.8  nm) or the superconducting coherence length (between 2 to   7 nm) as the third-dimension distance. Such choices will lead to different  magnitudes of $S^d_{sheet}$, which will  remain comparable to $k_B$ in order of magnitude. Nevertheless, in all these cases, the difference between compounds appears to be larger than the experimental uncertainty (for a recent discussion of the cuprate data, following ref. \cite{Rischau2021}, see \cite{HUEBENER20211353975}).

\begin{table}
 
\begin{tabular}{|cccccc|}
\hline
\centering
Compound & $T_{c}$ & N$^{peak}$&  $\rho^{peak}$ &  c & S$_{d}^{sheet}$ \\
& [K] & [$\mu$/K] & [$\mu \Omega$ cm]  & [nm] &  [10$^{-23}$J/K]  \\
\hline
YBa$_2$Cu$_3$O$_{7-\delta}$ (12 T)\cite{Li1994} & 92 & 2.6  &  50 & 1.19 & 1.3 \\
Bi$_{2}$Sr$_{2}$Ca Cu$_2$$_2$O$_{8+x}$ (12 T) \cite{Li1994}& 95 & 4.1 & 50 & 3.1 & 5.1\\
Nd$_{2-x}$Ce$_x$CuO$_4$ (1 T) \cite{Gollnik1998}& 24 & 1.6 & 20 &  1.2 & 1.9 \\
 La$_{1.92}$Sr$_{0.08}$CuO$_4$ (12 T) \cite{Capan2002}& 29 & 9.1 & 260 & 1.2 & 0.88  \\
\hline
\end{tabular}
\caption{\textbf{Extracting vortex entropy per sheet in cuprates-}The maximum Nernst signal, the  resistivity corresponding to temperature and magnetic field of the Nernst peak, the c-axis lattice parameter and the extracted entropy per vortex per sheet in the three cuprate superconductors.}
\label{Tab1}
\end{table}

\begin{figure*}
\centering
\includegraphics[width=16cm]{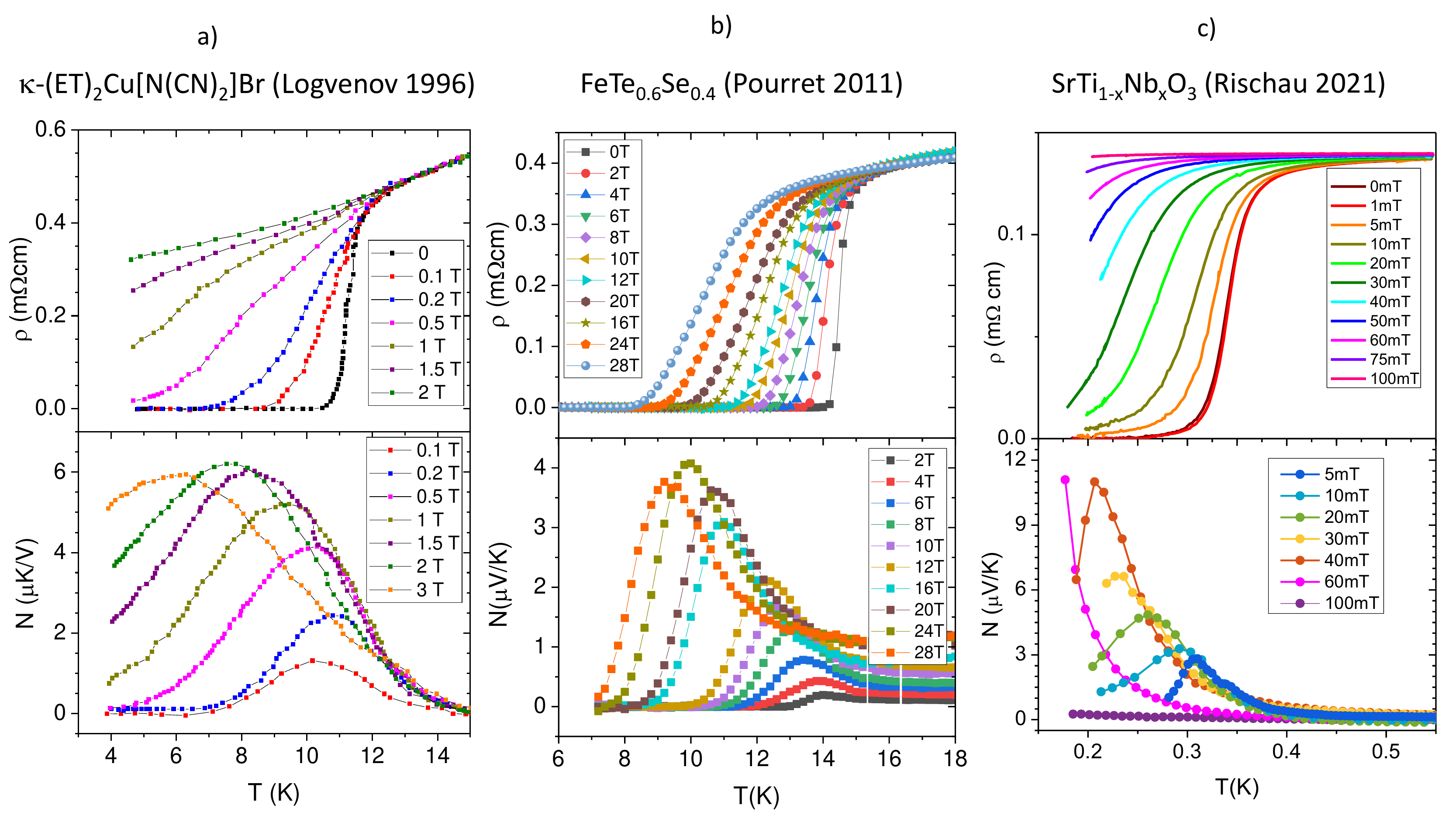}
\caption{\textbf{Nernst effect and resistivity in three superconductors:} Temperature dependence of resistivity and the Nernst signal in an organic superconductor \cite{Logvenov1996}, in an iron-based superconductor \cite{Pourret2011} and in strontium titanate \cite{Rischau2021}. In all cases, the Nernst signal peaks in the vortex liquid state at a a fraction of the critical temperature and a fraction of the zero-temperature upper critical field. This data has been used for values expressed in table \ref{Tab2}.}
\label{fig:three-super}
\end{figure*}
\textbf{Other uncommon superconductors} -  The Nernst effect in $\kappa$-(BEDT-TTF)$_2$X$_2$ family of organic superconductors was studied by two different groups ~\cite{Logvenov1996,Nam2007}. Logevenov \textit{et al.} ~\cite{Logvenov1996} also reported the resistivity of their samples. In the iron-based superconductor, FeTe$_{0.6}$Se$_{0.4}$, a Nernst peak in the vortex liquid state was detected by Pourret \textit{et al.} \cite{Pourret2011} who reported concomitant flux flow resistivity data, too.  Finally, the vortex Nernst signal and concomitant resistivity data were reported in the case of superconducting strontium titanate by Rischau \textit{et al.} \cite{Rischau2021}. These three data sets are reproduced in Fig.\ref{fig:three-super} and summarized in table \ref{Tab2}. Here also, one may contest the choice of lattice parameter as the relevant distance to pass from 3D to 2D. Note that the c-axis superconducting coherence is more than one order of magnitude longer in cubic strontium titanate than in cuprates. Therefore the similarity would disappear if one takes it as the relevant perpendicular distance.

\begin{table}
\begin{tabular}{|cccccc|}
\hline
\centering
Compound & $T_{c}$ & N$^{peak}$&  $\rho^{peak}$ &  c & S$_{d}^{sheet}$ \\
& [K] & [$\mu$/K] & [$\mu \Omega$ cm]  & [nm] &  [10$^{-23}$J/K]  \\
\hline
$\kappa$-(ET)$_2$Cu[N(CN)$_2$]Br (2 T)\cite{Logvenov1996} & 11 & 6.1  &  380 & 2.9 & 0.96\\
FeTe$_{0.6}$Se$_{0.4}$ (24 T) \cite{Pourret2011}& 14 & 4 & 48 & 0.58 & 0.96\\
SrTi$_{1-x}$Nb$_x$O$_3$ (0.06 T) \cite{Rischau2021}& 0.35 & 11 & 100 &  0.39 & 0.89 \\

\hline
\end{tabular}
\caption{\textbf{Extracting vortex entropy per sheet in three other superconductors-}The maximum Nernst signal, the  resistivity corresponding to temperature and magnetic field of the Nernst peak, the c-axis lattice parameter and the extracted entropy per vortex per sheet in the three superconductors. }
\label{Tab2}
\end{table}
\textbf{Historical Nernst data on superconducting niobium and its alloys} - We saw that in many superconductors, explored during the last few decades, the amplitude of $\alpha_{xy}$ leads to extracting an entropy per vortex per sheet of the order of $k_B$. However, as one can see in tables \ref{Tab1}, and \ref{Tab2},  the resistivity of the normal state in these superconductors is relatively large. One may wonder if the observation holds for superconductors with a higher conductivity. 

Most historical explorations of the Nernst effect did not explicitly report  the measured amplitude of S$_{xy}$. They focused on the measured transverse voltage and often extracted S$^d$.

In the case of niobium, Huebener and Seher \cite{Huebener1969} studied high-purity foils and found a Nernst signal as large as S$_{xy}\sim \mu V/K$ in samples with a residual resistivity as low as 20 n$\Omega$cm. This would imply an entropy per vortex per sheet exceeding $k_B$ by several orders of magnitude. 

In another study, de Lange and Otter \cite{deLange1975} quantified the magnitude of the vortex transport entropy,  S$_d$ in a niobium alloy (Nb$_{80}$Mo$_{20}$) and found that it peaks to $S_d\simeq 10^{-11} J/K.m$.  Combined with a lattice parameter of 0.3 nm, this yields a sheet entropy, S$_{d}^{sheet}$ exceeding 200 $k_B$.

Thus, niobium differs from the superconductors listed in the two tables. This suggests that S$_{d}^{sheet}\approx$ $k_B$ is a lower bound. On the other hand, in order of magnitude, the amplitude of S$_{xy}$ in this low-resistivity superconductor is comparable  to those listed above.

\subsection{Theory of the vortex transport entropy}
As mentioned in the introduction, as early as its discovery, the vortex Nernst effect was a subject of theoretical investigations \cite{Stephen1966,Clem1968,Maki1968}. In 2010, Sergeev and co-workers~\cite{Sergeev_2010}, reviewing these earlier theories  argued that magnetization currents do not transfer the thermal energy and therefore, the vortex transport entropy is much smaller than what was believed to be. 

Sergeev \textit{et al.}~\cite{Sergeev_2010} contested the validity of the following expression for the vortex transport entropy derived by Stephen~\cite{Stephen1966}:

\begin{equation}
S^{EM}_{d} = -\frac{\phi_0}{4\pi}\frac{\partial H_{c1}}{\partial T}
\label{s1}
\end{equation}

This expression assumes that the entropy of the vortex is set by the temperature derivative of the energy cost to introduce a vortex at the lower critical field, $H_{c1}$. But, according to Sergeev \textit{et al.}~\cite{Sergeev_2010}, such an assumption implies that supercurrents transport entropy. They derived the following expression for the vortex transport entropy : 
\begin{equation}
S^{core}_{d} \simeq -\pi \xi^2 \frac{\partial }{\partial T}\frac{H_c^2}{8\pi}
\label{s2}
\end{equation}

This theoretical expression for S$^d$ failed to explain the case of a BCS superconductor with a vortex liquid regime and well-established material-dependent parameters. Indeed, SrTi$_{1-x}$Nb$_x$O$_3$ at optimal doping (x=0.01) is such a superconductor. This is a s-wave superconductor \cite{Lin2014_multiple,Lin2015} where both the lower and the upper critical fields have been experimentally measured~\cite{Collignon2017} ($H_{c1}$(0)= 4.8 Oe; $H_{c2}$(0)= 240. Inserting these numbers in Eq. \ref{s2}, one finds $S^{EM}_d \approx  5.2 \times 10^{-12}$J/K. m. 

However, the experimental  $S_{d}$ is $\approx  2.3 \times 10^{-14}$J/K. m (equal to the ratio of $S^d_{sheet}$ and c in Table \ref{Tab2}). Thus, there is a fifty-fold discrepancy between theory and experiment.

Compared to  Eq. \ref{s1}, Eq. \ref{s2} leads to a downward revision to the amount of entropy carried by a vortex \cite{Sergeev_2010}. Nevertheless, what it yields is still more than one order of magnitude larger than what the the experiment finds. This discrepancy, together with the empirical observation that S$^d_{sheet}\approx k_B$, motivates us to explore other reasons for a theoretical overestimation of vortex transport entropy. This is the subject of the next section \footnote{For alternative proposals following the experimental observation reported in ref. \cite{Rischau2021}, see \cite{Segeev2021} and \cite{diamantini2022}.}.

\section{Distinguishing between vortex entropy at rest and in motion}

The experimental Nernst data and the  theoretical expectation (Eq. \ref{s2}) are confronted following a chain of reasoning, which contains a contestable assumption: The  entropy stocked in the core remains intact when transported by a mobile vortex.

In general, it is true that the magnitude of $\alpha_{xy}$quantifies the ratio of entropy to the magnetic flux of a mobile carrier \cite{Behnia2016} (See also \cite{Bergman2010}).  When the two opposing forces exerted on the carrier (the thermal force on its entropy and the Lorentz force on its magnetic flux) cancel each other, there is a finite $\alpha_{xy}$. It represents the ratio of a charge density flow to a perpendicular thermal gradient. This picture gives an account of the expressions for $\alpha_{xy}$ generated by electronic quasi-particles or fluctuating Cooper pairs, given the amount of entropy or magnetic flux they carry with themselves \cite{Behnia2016}. 

Now, a superconducting vortex is bound to a magnetic flux and its core has an excess of entropy. Therefore, one would naively expect to see that its generate a finite $\alpha_{xy}$ corresponding to the ratio of the entropy inside the core to the quantum of magnetic flux. However, this overlooks the fact that in this mesoscopic entity, the entropy and the magnetic flux are not necessarily bound to each other.

\begin{figure*}
\centering
\includegraphics[width=16cm]{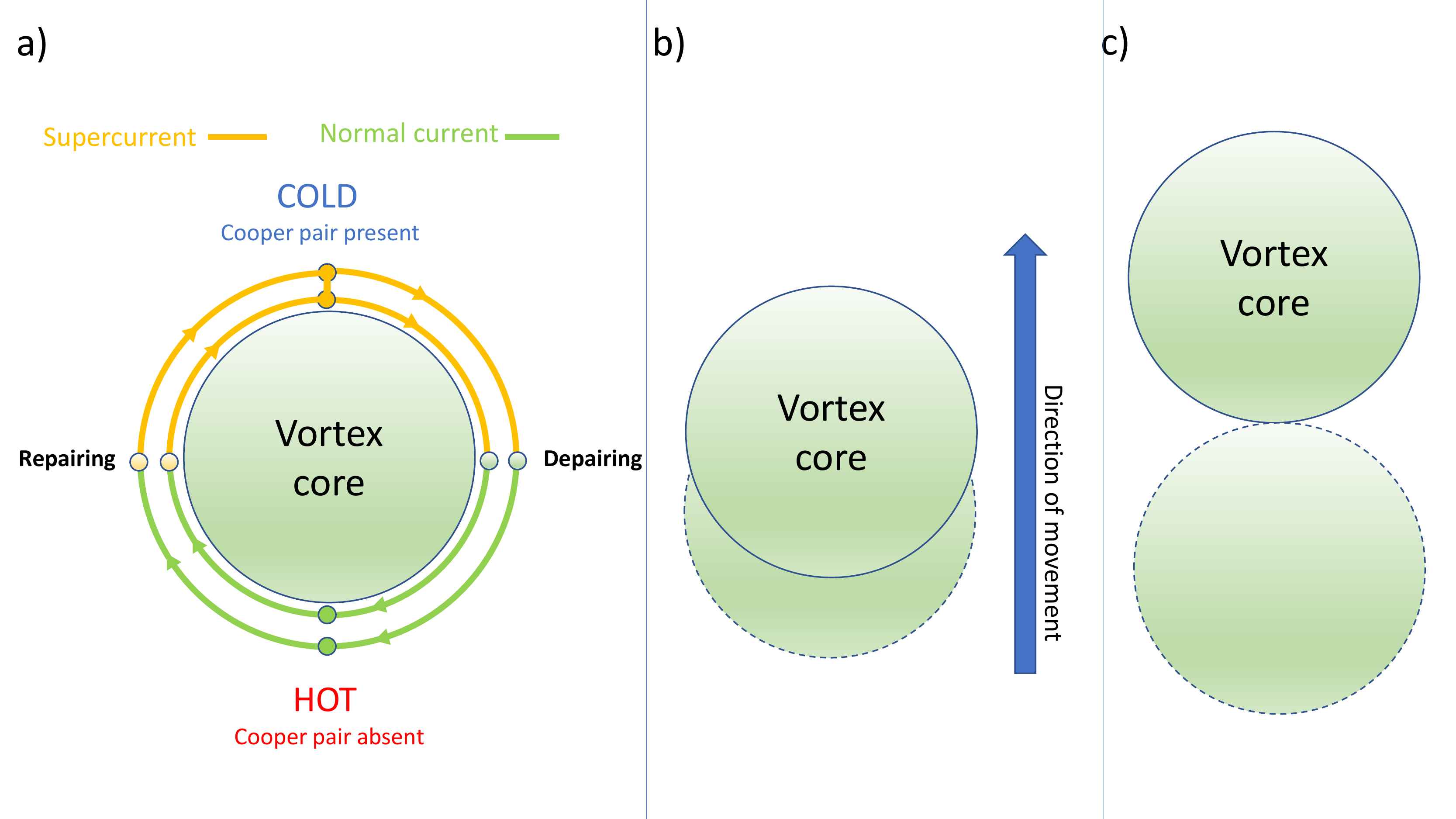}
\caption{\textbf{Vortex core and normal fluid in presence of thermal gradient:} a) A cooper pair in the immediate vicinity of a vortex core breaks up on the hot side of the vortex. b) As the vortex drifts under the influence of a thermal gradient, quasi-particles inside the core mix with those outside. c) When the vortex moves a distance of the order of its core radius, it can be totally stripped off its stocked entropy as a consequence of the exchange of inside and outside quasi-particles.}
\label{fig:core}
\end{figure*}
Fig.\ref{fig:core}a is a sketch of a superconducting vortex in presence of a thermal gradient. Around the vortex core, Cooper pairs whirl around the magnetic field with a velocity which increases as the core is approached. The core begins when this velocity becomes unsustainable and the pair breaks up. In presence of a thermal gradient, because of the temperature dependence of the superconducting gap, the depairing is favored on the hot side. A vortex drifting from the hot side to the cold side of the sample has thus a leaking tail. There is a steady transformation of peripheral Cooper pairs to core quasi-particles. Moreover, quasi-particles are not restricted to the normal core. At  fine temperature, a fraction of the total electronic density outside the vortex core is the normal fluid, which also diffuses heat in response to the thermal gradient.

The failure of Eq.\ref{s2} can be traced to the fact that because of the slow drift of the vortices, the mobile magnetic flux will be stripped of its entropy during a distance of the order of the core diameter (i.e. the superconducting coherence length, $\xi$) (See Fig.\ref{fig:core}b, c). 

Let us make a rough estimate of how slow a vortex should move in order to be stripped of its stocked entropy. The first relevant quantity is the tunneling rate between inside and outside of the vortex core. This rate is $\approx \frac{\Delta^2}{\hbar E_F}$.  The second is the number of quasi-particles stocked inside the core. In the most naive approach, the vortex core, can be assimilated to a normal-state cylinder of radius $\xi$ and the number of quasi-particles will be $\approx k_F^2 \xi^2$. In a more sophisticated approach, what the core contains are $\approx k_F \xi$ Andreev bound states \cite{Kopnin1991,Stone1996}. Therefore, one can estimate the time it takes to entirely reconfigure the vortex to be of the order of $k_F \xi \times \frac{ \hbar E_F}{\Delta^2}$. 

A drifting vortex with a velocity of $v_L$ will see its core content replaced if the drift time along a distance of $\xi$  is longer than this time. This will happen if :

\begin{equation}
\frac{\xi}{v_L} > k_F \xi \times \frac{ \hbar E_F}{\Delta^2}
\label{s3}
\end{equation}

Neglecting numerical factors of the order of unity, one finds that the inequality \ref{s3} eventually leads to: 
\begin{equation}
\frac{v_L}{v_F} < (\frac{\Delta}{E_F})^2
\label{s4}
\end{equation}

Numerical prefactors of the order of unity have been neglected. Therefore the vortex core will be stripped off its quasi-particles and entropy if the drift velocity  is much smaller  than the Fermi velocity, v$_F$, multiplied by the square of the ratio of the superconducting gap to the Fermi energy $\frac{\Delta}{E_F}$.  

Now, let us consider the numbers. The measured electric field in a Nernst experiment is of the order of $\approx 10^{-3}$Vm$^{-1}$ in a magnetic field of $\approx 1 $ T Therefore, the flux lines drift with a velocity of $10^{-3}$ m/s. The Fermi velocity, on the other hand, is of the order of $10^5$ m/s. This is a discrepancy of 8 orders of magnitude. It allows inequality \ref{s4} to hold, provided that $\frac{\Delta}{E_F}$ does not fall below a very small number. Interestingly, all superconductors listed in tables \ref{Tab1} and \ref{Tab2} have relatively large $\frac{\Delta}{E_F}$ ratios (in the range of $10^{-1}$ to $10^{-2}$). On the other hand, in an elemental superconductor such as Nb, $\frac{\Delta}{E_F}$ can be as small as $10^{-4}$. In the latter case, the vortex may not be totally stripped from the entropy stocked in their core.

\subsection{An irreducible entropy per vortex ?}
Is there a lower boundary to the entropy of a moving vortex?  If the whole content of the vortex core get replaced during a movement as short as $\xi$, why should there be any mobile entropy at all?

In search of a possible answer to this question, let us recall what Volovik \cite{Volovik2003} pointed out about vortices in fermionic superfluids. He stated that analogs of the black hole horizon can occur in `\textit{liquids moving with velocities exceeding the local maximum attainable speed of quasiparticles. Then an inner observer, who uses only quasiparticles as a means of transferring the information, finds that some regions of space are not accessible for observation. For this observer, who lives in the quantum liquid, these regions are black holes.}'\cite{Volovik2003}

It is tempting to speculate that a lower boundary of $k_B$ln2 arises is a consequence of the event horizon surrounding a vortex. An information barrier between observers inside and outside would ensure the survival of this last bit of information when the vortex constituents are totally replaced.

\section{Concluding remarks} The transport properties of the vortex liquids are often more complex than what it seems at the first sight. This is true not only for the Vortex Nernst response, but also for flux flow resistivity \cite{Narayan_2003} and the Hall response in the vortex liquid regime \cite{Auerbach2020}.

A satisfactory theory of the Nernst response in the vortex liquid is yet to be formulated. This is in contrast with the case of the Nernst signal in the normal state of a superconductor, where the theory has been tested by multiple experimental studies \cite{Behnia2016,jotzu2021}. Here, I argued that the rough amplitude of the Nernst signal can be understood provided that the vortex liquid respects a bound to $\frac{\eta}{s}$ (the viscosity-to entropy density ratio), which is observed by other liquids. Moreover, the entropy inside a vortex core does not remain bound to the magnetic flux line during the drift if the ratio of the superconducting gap to the Fermi energy is above reasonable threshold. This may explain why the entropy per vortex is much smaller in superconductors with a larger $\frac{\Delta}{E_F}$ ratio, but not in superconducting niobium. 

The subject deserves further theoretical and experimental investigations.

\section{Acknowlegements}
I am grateful to   Beno\^it Fauqu\'e,  M. V. Feigel'man, S. A. Hartnoll, A. Kapitulnik, S. A. Kivelson, K. Trachenko, C.W. Rischau, and G. E. Volovik for discussions.  This work was supported by the Agence Nationale de la Recherche  (ANR-18-CE92-0020-01; ANR-19-CE30-0014-04).
\newline
\bibliographystyle{unsrt}
\bibliography{biblio.bib}

\begin{thebibliography}{10}

\bibitem{Behnia2016}
Kamran Behnia and Herv\'e Aubin.
\newblock Nernst effect in metals and superconductors: a review of concepts and
  experiments.
\newblock {\em Rep. Prog. Phys.}, 79(4):046502, 2016.

\bibitem{Galev1981}
V.~N. {Galev}, V.~A. {Kozlov}, N.~V. {Kolomoets}, S.~Ya. {Skipidarov}, and
  N.~A. {Tsvetkova}.
\newblock Exponential temperature dependence of the coefficient of the
  {Nernst-Ettingshausen} transverse effect in bismuth.
\newblock {\em Soviet Journal of Experimental and Theoretical Physics Letters},
  33:106, January 1981.

\bibitem{LOWELL1967}
J.~Lowell, J.S. Muñoz, and J.~Sousa.
\newblock Thermally induced voltages in the mixed state of type {II}
  superconductors.
\newblock {\em Physics Letters A}, 24(7):376--377, 1967.

\bibitem{solomon1967}
P.~R. Solomon and F.~A. Otter.
\newblock Thermomagnetic effects in superconductors.
\newblock {\em Phys. Rev.}, 164:608--618, Dec 1967.

\bibitem{HUEBENER1967947}
R.P. Huebener.
\newblock Thermal force on vortices in superconducting lead films.
\newblock {\em Solid State Communications}, 5(12):947--950, 1967.

\bibitem{Stephen1966}
M.~J. Stephen.
\newblock Galvanomagnetic and related effects in {Type-II} superconductors.
\newblock {\em Phys. Rev. Lett.}, 16:801--803, May 1966.

\bibitem{Clem1968}
John~R. Clem.
\newblock Motion of normal regions along a superconducting strip.
\newblock {\em Phys. Rev.}, 176:531--537, Dec 1968.

\bibitem{Maki1968}
Kazumi Maki.
\newblock Thermomagnetic effects in dirty type-{II} superconductors.
\newblock {\em Phys. Rev. Lett.}, 21:1755--1757, Dec 1968.

\bibitem{Huebener_1995}
R~P Huebener.
\newblock Superconductors in a temperature gradient.
\newblock {\em Superconductor Science and Technology}, 8(4):189--198, apr 1995.

\bibitem{Blatter1994}
G.~Blatter, M.~V. Feigel'man, V.~B. Geshkenbein, A.~I. Larkin, and V.~M.
  Vinokur.
\newblock Vortices in high-temperature superconductors.
\newblock {\em Rev. Mod. Phys.}, 66:1125--1388, Oct 1994.

\bibitem{Rischau2021}
Carl~Willem Rischau, Yuke Li, Beno\^{\i}t Fauqu\'e, Hisashi Inoue, Minu Kim,
  Christopher Bell, Harold~Y. Hwang, Aharon Kapitulnik, and Kamran Behnia.
\newblock Universal bound to the amplitude of the vortex {Nernst} signal in
  superconductors.
\newblock {\em Phys. Rev. Lett.}, 126:077001, Feb 2021.

\bibitem{Ussishkin}
Iddo Ussishkin, S.~L. Sondhi, and David~A. Huse.
\newblock Gaussian superconducting fluctuations, thermal transport, and the
  {Nernst} effect.
\newblock {\em Phys. Rev. Lett.}, 89:287001, Dec 2002.

\bibitem{Serbyn2009}
M.~N. Serbyn, M.~A. Skvortsov, A.~A. Varlamov, and Victor Galitski.
\newblock Giant {Nernst} effect due to fluctuating {Cooper} pairs in
  superconductors.
\newblock {\em Phys. Rev. Lett.}, 102:067001, Feb 2009.

\bibitem{Michaeli2009}
K.~Michaeli and A.~M. Finkelstein.
\newblock Fluctuations of the superconducting order parameter as an origin of
  the {Nernst} effect.
\newblock {\em {EPL} (Europhysics Letters)}, 86(2):27007, apr 2009.

\bibitem{Levchenko2011}
Alex Levchenko, M.~R. Norman, and A.~A. Varlamov.
\newblock Nernst effect from fluctuating pairs in the pseudogap phase of the
  cuprates.
\newblock {\em Phys. Rev. B}, 83:020506, Jan 2011.

\bibitem{Pourret2006}
A.~Pourret, H.~Aubin, J.~Lesueur, C.~A. Marrache-Kikuchi, L.~Berg{\'e},
  L.~Dumoulin, and K.~Behnia.
\newblock Observation of the {Nernst} signal generated by fluctuating {Cooper}
  pairs.
\newblock {\em Nature Physics}, 2(10):683--686, 2006.

\bibitem{Pourret2007}
A.~Pourret, H.~Aubin, J.~Lesueur, C.~A. Marrache-Kikuchi, L.~Berg\'e,
  L.~Dumoulin, and K.~Behnia.
\newblock {Length scale for the superconducting Nernst signal above ${T}_{c}$
  in ${\mathrm{Nb}}_{0.15}{\mathrm{Si}}_{0.85}$}.
\newblock {\em Phys. Rev. B}, 76:214504, Dec 2007.

\bibitem{Spathis2008}
P.~Spathis, H.~Aubin, A.~Pourret, and K.~Behnia.
\newblock Nernst effect in the phase-fluctuating superconductor {InO$_x$}.
\newblock {\em {EPL} (Europhysics Letters)}, 83(5):57005, sep 2008.

\bibitem{kokanovic2009}
I.~Kokanovi\ifmmode~\acute{c}\else \'{c}\fi{}, J.~R. Cooper, and M.~Matusiak.
\newblock Nernst effect measurements of epitaxial
  {${\mathbf{Y}}_{0.95}{\mathrm{Ca}}_{0.05}{\mathrm{Ba}}_{2}({\mathrm{Cu}}_{1\ensuremath{-}x}{\mathrm{Zn}}_{x}{)}_{3}{\mathbf{O}}_{y}$}
  and
  {${\mathbf{Y}}_{0.9}{\mathrm{Ca}}_{0.1}{\mathrm{Ba}}_{2}{\mathrm{Cu}}_{3}{\mathbf{O}}_{y}$}
  superconducting films.
\newblock {\em Phys. Rev. Lett.}, 102:187002, May 2009.

\bibitem{Chang2012}
J.~Chang, N.~Doiron-Leyraud, O.~Cyr-Choinière, G.~Grissonnanche,
  F.~Laliberté, E.~Hassinger, J-Ph Reid, R.~Daou, S.~Pyon, T.~Takayama,
  H.~Takagi, and L.~Taillefer.
\newblock Decrease of upper critical field with underdoping in cuprate
  superconductors.
\newblock {\em Nature Phys}, 8:751--756, 2012.

\bibitem{Tafti2014}
F.~F. Tafti, F.~Lalibert\'e, M.~Dion, J.~Gaudet, P.~Fournier, and Louis
  Taillefer.
\newblock {Nernst} effect in the electron-doped cuprate superconductor
  {${\mathrm{Pr}}_{2\ensuremath{-}x}{\mathrm{Ce}}_{x}{\mathrm{CuO}}_{4}$}:
  Superconducting fluctuations, upper critical field ${H}_{c2}$, and the origin
  of the ${T}_{c}$ dome.
\newblock {\em Phys. Rev. B}, 90:024519, Jul 2014.

\bibitem{Pourret2009}
A~Pourret, P~Spathis, H~Aubin, and K~Behnia.
\newblock Nernst effect as a probe of superconducting fluctuations in
  disordered thin films.
\newblock {\em New Journal of Physics}, 11(5):055071, May 2009.

\bibitem{Cyr2018}
O.~Cyr-Choini\`ere, R.~Daou, F.~Lalibert\'e, C.~Collignon, S.~Badoux,
  D.~LeBoeuf, J.~Chang, B.~J. Ramshaw, D.~A. Bonn, W.~N. Hardy, R.~Liang, J.-Q.
  Yan, J.-G. Cheng, J.-S. Zhou, J.~B. Goodenough, S.~Pyon, T.~Takayama,
  H.~Takagi, N.~Doiron-Leyraud, and Louis Taillefer.
\newblock Pseudogap temperature ${T}^{*}$ of cuprate superconductors from the
  {Nernst} effect.
\newblock {\em Phys. Rev. B}, 97:064502, Feb 2018.

\bibitem{Xu2000}
Z.~A. Xu, N.~P. Ong, Y.~Wang, T.~Kakeshita, and S.~Uchida.
\newblock Vortex-like excitations and the onset of superconducting phase
  fluctuation in underdoped {La$_{2-x}$Sr$_x$CuO$_4$}.
\newblock {\em Nature}, 406(6795):486--488, Aug 2000.

\bibitem{Xu2020}
Liangcai Xu, Xiaokang Li, Linchao Ding, Taishi Chen, Akito Sakai, Beno\^{\i}t
  Fauqu\'e, Satoru Nakatsuji, Zengwei Zhu, and Kamran Behnia.
\newblock Anomalous transverse response of {$Co_{2}MnGa$} and universality of
  the room-temperature
  ${\ensuremath{\alpha}}_{ij}^{A}/{\ensuremath{\sigma}}_{ij}^{A}$ ratio across
  topological magnets.
\newblock {\em Phys. Rev. B}, 101:180404, May 2020.

\bibitem{Behnia2009}
Kamran Behnia.
\newblock {The Nernst effect and the boundaries of the Fermi liquid picture}.
\newblock {\em Journal of Physics: Condensed Matter}, 21(11):113101, 2009.

\bibitem{Sugihara}
Ko~Sugihara.
\newblock Thermomagnetic effects in bismuth. ii. {Nernst-Ettingshausen} effect.
\newblock {\em Journal of the Physical Society of Japan}, 27(2):362--370, 1969.

\bibitem{Mangez1976}
J.~H. Mangez, J.~P. Issi, and J.~Heremans.
\newblock Transport properties of bismuth in quantizing magnetic fields.
\newblock {\em Phys. Rev. B}, 14:4381--4385, Nov 1976.

\bibitem{Behnia2007}
Kamran Behnia, Marie-Aude M\'easson, and Yakov Kopelevich.
\newblock Nernst effect in semimetals: The effective mass and the figure of
  merit.
\newblock {\em Phys. Rev. Lett.}, 98:076603, Feb 2007.

\bibitem{Xu2008}
X.~F. Xu, Z.~A. Xu, T.~J. Liu, D.~Fobes, Z.~Q. Mao, J.~L. Luo, and Y.~Liu.
\newblock Band-dependent normal-state coherence in {Sr$_2$RuO$_{4}$}: Evidence
  from nernst effect and thermopower measurements.
\newblock {\em Phys. Rev. Lett.}, 101:057002, Jul 2008.

\bibitem{Jin2021}
Hao Jin, Alessandro Narduzzo, Minoru Nohara, Hidenori Takagi, N.~E. Hussey, and
  Kamran Behnia.
\newblock Positive seebeck coefficient in highly doped
  {La$_{1-x}$Sr$_x$CuO$_4$}(x=0.3); its origin and implication.
\newblock {\em Journal of the Physical Society of Japan}, 90(5):053702, 2021.

\bibitem{falkovich2011fluid}
G.~Falkovich.
\newblock {\em Fluid Mechanics: A Short Course for Physicists}.
\newblock Cambridge University Press, 2011.

\bibitem{Purcell}
E.~M. Purcell.
\newblock Life at low reynolds number.
\newblock {\em American Journal of Physics}, 45(1):3--11, 1977.

\bibitem{Trachenko2021}
Kostya Trachenko and Vadim~V. Brazhkin.
\newblock The quantum mechanics of viscosity.
\newblock {\em Physics Today}, 74(12):66--67, 2021.

\bibitem{Trachenko2020}
K.~Trachenko and V.~V. Brazhkin.
\newblock Minimal quantum viscosity from fundamental physical constants.
\newblock {\em Science Advances}, 6(17), 2020.

\bibitem{Trachenko2021c}
K.~Trachenko, M.~Baggioli, K.~Behnia, and V.~V. Brazhkin.
\newblock Universal lower bounds on energy and momentum diffusion in liquids.
\newblock {\em Phys. Rev. B}, 103:014311, Jan 2021.

\bibitem{Cremonini2011}
Sera Cremonini.
\newblock The shear viscosity to entropy ratio: A status report.
\newblock {\em Modern Physics Letters B}, 25(23):1867--1888, 2011.

\bibitem{Cremonini2012}
Sera Cremonini, Umut G{\"u}rsoy, and Phillip Szepietowski.
\newblock On the temperature dependence of the shear viscosity and holography.
\newblock {\em Journal of High Energy Physics}, 2012(8):167, Aug 2012.

\bibitem{Policastro2001}
G.~Policastro, D.~T. Son, and A.~O. Starinets.
\newblock {Shear Viscosity of Strongly Coupled
  $N\phantom{\rule{0ex}{0ex}}=\phantom{\rule{0ex}{0ex}}4$ Supersymmetric
  Yang-Mills Plasma}.
\newblock {\em Phys. Rev. Lett.}, 87:081601, Aug 2001.

\bibitem{Kovtun2005}
P.~K. Kovtun, D.~T. Son, and A.~O. Starinets.
\newblock Viscosity in strongly interacting quantum field theories from black
  hole physics.
\newblock {\em Phys. Rev. Lett.}, 94:111601, Mar 2005.

\bibitem{nist}
{\em National Institute of Standards and Technology database, see
  https://webbook.nist.gov/chemistry/fluid}, 2020.

\bibitem{Gollnik1998}
F.~Gollnik and M.~Naito.
\newblock Doping dependence of normal- and superconducting-state transport
  properties of
  {${\mathrm{Nd}}_{2\ensuremath{-}x}{\mathrm{Ce}}_{x}{\mathrm{CuO}}_{4\ifmmode\pm\else\textpm\fi{}y}$}
  thin films.
\newblock {\em Phys. Rev. B}, 58:11734--11752, Nov 1998.

\bibitem{Li1994}
H.-C. Ri, R.~Gross, F.~Gollnik, A.~Beck, R.~P. Huebener, P.~Wagner, and
  H.~Adrian.
\newblock {Nernst, Seebeck, and Hall} effects in the mixed state of
  {${\mathrm{YBa}}_{2}$${\mathrm{Cu}}_{3}$${\mathrm{O}}_{7\mathrm{\ensuremath{-}}\mathrm{\ensuremath{\delta}}}$}
  and
  {${\mathrm{Bi}}_{2}$${\mathrm{Sr}}_{2}$${\mathrm{CaCu}}_{2}$${\mathrm{O}}_{8+\mathit{x}}$}
  thin films: A comparative study.
\newblock {\em Phys. Rev. B}, 50:3312--3329, Aug 1994.

\bibitem{Capan2002}
C.~Capan, K.~Behnia, J.~Hinderer, A.~G.~M. Jansen, W.~Lang, C.~Marcenat,
  C.~Marin, and J.~Flouquet.
\newblock Entropy of vortex cores near the superconductor-insulator transition
  in an underdoped cuprate.
\newblock {\em Phys. Rev. Lett.}, 88:056601, Jan 2002.

\bibitem{Huebener1979}
R.~P. Huebener.
\newblock {\em Magnetic Flux Structures in Superconductors}.
\newblock Springer-Verlag, Berlin, 1979.

\bibitem{Palstra1990}
T.~T.~M. Palstra, B.~Batlogg, L.~F. Schneemeyer, and J.~V. Waszczak.
\newblock Transport entropy of vortex motion in
  {${\mathrm{YBa}}_{2}$${\mathrm{Cu}}_{3}$${\mathrm{O}}_{7}$}.
\newblock {\em Phys. Rev. Lett.}, 64:3090--3093, Jun 1990.

\bibitem{Wang2001}
Yayu Wang, Z.~A. Xu, T.~Kakeshita, S.~Uchida, S.~Ono, Yoichi Ando, and N.~P.
  Ong.
\newblock Onset of the vortexlike {Nernst} signal above ${T}_{c}$ in
  {${\mathrm{La}}_{2\ensuremath{-}x}{\mathrm{Sr}}_{x}{\mathrm{CuO}}_{4}$} and
  {${\mathrm{Bi}}_{2}{\mathrm{Sr}}_{2\ensuremath{-}y}{\mathrm{La}}_{y}{\mathrm{CuO}}_{6}$}.
\newblock {\em Phys. Rev. B}, 64:224519, Nov 2001.

\bibitem{Wang2002}
Yayu Wang, N.~P. Ong, Z.~A. Xu, T.~Kakeshita, S.~Uchida, D.~A. Bonn, R.~Liang,
  and W.~N. Hardy.
\newblock High field phase diagram of cuprates derived from the {Nernst}
  effect.
\newblock {\em Phys. Rev. Lett.}, 88:257003, Jun 2002.

\bibitem{Wang2006}
Yayu Wang, Lu~Li, and N.~P. Ong.
\newblock Nernst effect in high-${T}_{c}$ superconductors.
\newblock {\em Phys. Rev. B}, 73:024510, Jan 2006.

\bibitem{Rullier2006}
F.~Rullier-Albenque, R.~Tourbot, H.~Alloul, P.~Lejay, D.~Colson, and A.~Forget.
\newblock Nernst effect and disorder in the normal state of high-${T}_{c}$
  cuprates.
\newblock {\em Phys. Rev. Lett.}, 96:067002, Feb 2006.

\bibitem{Balci2003}
Hamza Balci, C.~P. Hill, M.~M. Qazilbash, and R.~L. Greene.
\newblock Nernst effect in electron-doped
  {${\mathrm{Pr}}_{2\ensuremath{-}x}{\mathrm{Ce}}_{x}{\mathrm{CuO}}_{4}$}.
\newblock {\em Phys. Rev. B}, 68:054520, Aug 2003.

\bibitem{HUEBENER20211353975}
R.P. Huebener and H.-C. Ri.
\newblock Vortex transport entropy in cuprate superconductors and {Boltzmann}
  constant.
\newblock {\em Physica C: Superconductivity and its Applications}, 591:1353975,
  2021.

\bibitem{Logvenov1996}
G.Yu. Logvenov, H.~Ito, T.~Ishiguro, G.~Saito, S.~Takasaki, J.~Yamada, and
  H.~Anzai.
\newblock Anomalous nernst effect in the mixed state of the two-band organic
  superconductors {$\kappa$-(BEDT-TTF)$_2$Cu[N(CN)$_2$]Br} and
  {$\kappa$-(BEDT-TTF)2Cu(NCS)$_2$}.
\newblock {\em Physica C: Superconductivity}, 264(3):261 -- 267, 1996.

\bibitem{Pourret2011}
Alexandre Pourret, Liam Malone, Arlei~B. Antunes, C.~S. Yadav, P.~L. Paulose,
  Beno\^{\i}t Fauqu\'e, and Kamran Behnia.
\newblock Strong correlation and low carrier density in
  {${\mathrm{Fe}}_{1+y}{\mathrm{Te}}_{0.6}{\mathrm{Se}}_{0.4}$} as seen from
  its thermoelectric response.
\newblock {\em Phys. Rev. B}, 83:020504, Jan 2011.

\bibitem{Nam2007}
Moon-Sun Nam, Arzhang Ardavan, Stephen~J. Blundell, and John~A. Schlueter.
\newblock Fluctuating superconductivity in organic molecular metals close to
  the {Mott} transition.
\newblock {\em Nature}, 449(7162):584--587, 2007.

\bibitem{Huebener1969}
R.~P. Huebener and A.~Seher.
\newblock Nernst effect and flux flow in superconductors. i. niobium.
\newblock {\em Phys. Rev.}, 181:701--709, May 1969.

\bibitem{deLange1975}
O.~L. de~Lange and F.~A. Otter.
\newblock Flux flow effects in a nearly reversible type {II} superconductor.
\newblock {\em Journal of Low Temperature Physics}, 18(1):31--42, Jan 1975.

\bibitem{Sergeev_2010}
A.~Sergeev, M.~Reizer, and V.~Mitin.
\newblock Thermomagnetic vortex transport: Transport entropy revisited.
\newblock {\em {EPL} (Europhysics Letters)}, 92(2):27003, oct 2010.

\bibitem{Lin2014_multiple}
Xiao Lin, Adrien Gourgout, German Bridoux, Francois Jomard, Alexandre Pourret,
  Beno\^{\i}t Fauqu\'e, Dai Aoki, and Kamran Behnia.
\newblock {Multiple nodeless superconducting gaps in optimally doped
  ${\mathrm{SrTi}}_{1\ensuremath{-}x}{\mathrm{Nb}}_{x}{\mathrm{O}}_{3}$}.
\newblock {\em Phys. Rev. B}, 90:140508, Oct 2014.

\bibitem{Lin2015}
Xiao Lin, Carl~Willem Rischau, Cornelis~J. van~der Beek, Beno\^{\i}t Fauqu\'e,
  and Kamran Behnia.
\newblock {$s$-wave superconductivity in optimally doped
  SrTi$_{1-x}$Nb$_x$O$_3$ unveiled by electron irradiation}.
\newblock {\em Phys. Rev. B}, 92:174504, Nov 2015.

\bibitem{Collignon2017}
Cl\'ement Collignon, Beno\^{\i}t Fauqu\'e, Antonella Cavanna, Ulf Gennser,
  Dominique Mailly, and Kamran Behnia.
\newblock Superfluid density and carrier concentration across a superconducting
  dome: The case of strontium titanate.
\newblock {\em Phys. Rev. B}, 96:224506, Dec 2017.

\bibitem{Segeev2021}
Andrei Sergeev and Michael Reizer.
\newblock Entropy-based theory of thermomagnetic phenomena.
\newblock {\em International Journal of Modern Physics B}, 35(18):2150190,
  2021.

\bibitem{diamantini2022}
Maria~Cristina Diamantini, Carlo~A. Trugenberger, and Valerii~M. Vinokur.
\newblock Universal upper bound for the entropy of superconducting vortices and
  the quantum {Nernst} effect.
\newblock {\em Quantum Reports}, 4(1):16--21, 2022.

\bibitem{Bergman2010}
Doron~L. Bergman and Vadim Oganesyan.
\newblock Theory of dissipationless {Nernst} effects.
\newblock {\em Phys. Rev. Lett.}, 104:066601, Feb 2010.

\bibitem{Kopnin1991}
N.~B. Kopnin and M.~M. Salomaa.
\newblock Mutual friction in superfluid {$^{3}\mathrm{He}$}: Effects of bound
  states in the vortex core.
\newblock {\em Phys. Rev. B}, 44:9667--9677, Nov 1991.

\bibitem{Stone1996}
Michael Stone.
\newblock Spectral flow, {Magnus} force, and mutual friction via the geometric
  optics limit of {Andreev} reflection.
\newblock {\em Phys. Rev. B}, 54:13222--13229, Nov 1996.

\bibitem{Volovik2003}
Grigory~E. Volovik.
\newblock {\em {The Universe in a Helium Droplet}}.
\newblock Oxford University Press, 2003.

\bibitem{Narayan_2003}
Onuttom Narayan.
\newblock Driving force on an {Abrikosov} vortex.
\newblock {\em Journal of Physics A: Mathematical and General},
  36(23):L373--L377, may 2003.

\bibitem{Auerbach2020}
Assa Auerbach and Daniel~P. Arovas.
\newblock {Hall anomaly and moving vortex charge in layered superconductors}.
\newblock {\em SciPost Phys.}, 8:061, 2020.

\bibitem{jotzu2021}
Gregor Jotzu, Guido Meier, Alice Cantaluppi, Andrea Cavalleri, Daniele
  Pontiroli, Mauro Ricc{\`o}, Arzhang Ardavan, and Moon-Sun Nam.
\newblock Superconducting fluctuations observed far above {T$_c$} in the
  isotropic superconductor {K$_3$C$_{60}$}.
\newblock {\em arXiv preprint arXiv:2109.08679}, 2021.

\end{thebibliography}
\end{document}